\documentstyle[12pt]{article}

\date{}
\begin{document}
\makeatletter



\title{\bf  Theta Term in QCD sum rules at Finite Temperature and the Neutron Electric Dipole Moment }

\author { {\Large Mohamed Chabab} \\ \\ 
 {\it LPHEA, Physics  Department, Faculty of Science-Semlalia,}
\\  {\it P.O. Box 2390, Marrakesh, Morocco} 
\\{\it mchabab@ucam.ac.ma} }

\maketitle

\begin{abstract} 
By using thermal QCD sum rules to investigate the $\bar{\theta}$ induced
neutron electric dipole moment $d_n$, we have examined the behaviour of broken CP symmetry
 at finite temperature. We find that, below the critical temperature,  the ratio $\mid {d_n \over
\bar{\theta}}\mid$ slightly decreases but survives at temperature effects, implying T nonrestoration of
CP-invariance \cite{Chabab}.

\end{abstract}

\section*{INTRODUCTION}

To establish connection between particle physics and cosmology, it is essential to study the behaviour of symmetry breaking in the early universe, i.e. at high temperature. The CP symmetry is, without doubt, one of the most fundamental
symmetries in nature. It is
intimately related to theories  of interactions between elementary
particles and represents a cornerstone in constructing grand
unified and supersymmetric models. However, its breaking still carries a cloud of
mystery in particle physics and cosmology since it is necessary to explain
baryogenesis and is required by theories with domain walls.

According to the CPT theorem, CP violation implies T violation. The latter is
tested through the measurement of the neutron electric dipole
moment (NEDM) $d_n$. The upper experimental limit  gives confidence
that the NEDM can be another manifestation of CP breaking. 
In the Standard Model, CP violation is originated from two sources: the first source, which appears in the electroweak sector, is parametrized by a single phase in the Cabbibo-Kobayashi-Maskawa (CKM) quark mixing matrix \cite{CKM}. The other source
 is due to the the so called $\theta$-term of QCD. In fact, the  the QCD effective lagrangian contains an additional CP-odd four dimensional operator embedded in the following topological term:

\begin{equation}
L_{\theta}=\theta  {\alpha_s\over 8
\pi}G_{\mu\nu}\tilde{G}^{\mu\nu},
\end{equation}

where $G_{\mu\nu}$ is the gluonic field strength,
$\tilde{G}^{\mu\nu}$ is its dual and $\alpha_s$ is the strong
coupling constant. The  $G_{\mu\nu}\tilde{G}^{\mu\nu}$ quantity is
a total derivative, consequently it contribute to the physical
observables only through non perturbative effects. The NEDM is
related to the $\bar \theta$-angle by the following relation:

\begin{equation}
d_n\sim {e\over M_n}({m_q\over M_n})\bar \theta \sim \{
\begin{array}{c}
2.7\times 10^{-16}\overline{\theta }\qquad \cite{Baluni}\\
5.2\times 10^{-16}\overline{\theta }\qquad \cite{cvvw}
\end{array}
\end{equation}

and consequently, according to the experimental
measurements $d_n<1.1\times 10^{-25}ecm$ \cite{data}, the $\bar
\theta $ parameter must be less than $2\times10^{-10}$
\cite{peccei2}. The well known strong CP problem consists in
explaining the smallness of $\bar{\theta}$. In this regard,
several scenarios were suggested. The most popular one is the Peccei and Quinn \cite{PQ}, in which $\bar{\theta]}$
is identified to a very light pseudo goldstone boson called the axion. This particle arises from the
spontaneous breakdown of a global $U_A(1)$ symmetry and may well be
important to explain the puzzle of dark matter  providing a peace of information on 
the missing mass of the universe \cite{LS}.

Our aim in this work is to study the thermal behaviour of the CP symmetry breaking and the temperature effects on its restoration. 
This is motivated by the possibility to restore some broken symmetries by increasing the temperature.

This paper is organized as follows: Section 2 is devoted to the calculations of the NEDM
induced by the  $\bar {\theta}$ using QCD sum rules. In
section 3,  we show how one introduces temperature in QCD sum
rules calculations. The last section is devoted to a discussion and
qualitative analysis of the thermal effects on the CP symmetry.

\section*{ NEDM from QCD sum rules}

Since the NEDM has essaentially a non perturbative nature, one should any how take 
into account some effects which escape the perturbative treatement. One such approach, 
dealing with the strong strong coupling regime and based on first principales of the 
theory is QCD sum rules. Such approach has been applied successfully 
to the investigation of hadronic properties
at low energies, particularly to certain baryonic magnetic form factor. 
In order to derive the NEDM through through this approach \cite{PR,PR1}, we consider a
 lagrangian containing the following P and CP violating operators:
\begin{equation}
L_{P,CP}=-\theta_q m_* \sum_f \bar{q}_f i\gamma_5 q_f +\theta
{\alpha_s\over 8 \pi}G_{\mu\nu}\tilde{G}^{\mu\nu}.
\end{equation}
$\theta_q$ and $\theta$ are respectively two angles coming from
the chiral and the topological terms and $m_*$ is the quark
reduced mass given by $m_*$=$m_um_d \over{m_u +m_d} $. Then, as usual, we start
 from the two points correlation function in
QCD background with a nonvanishing $\theta$ and in the presence of
a constant external electomagnetic field $ F^{\mu\nu}$ \cite{PR1}:
\begin{equation}
\Pi(q^2) = i \int d^4x
e^{iqx}<0|T\{\eta(x)\bar{\eta}(0)\}|0>_{\theta,F} .
\end{equation}
where $\eta(x)$ is the neutron  interpolating current  \cite{I}:
\begin{equation}
\eta
=2\epsilon_{abc}\{(d^T_aC\gamma_5u_b)d_c+\beta(d^T_aCu_b)\gamma_5d_c\},
\end{equation}
and  $\beta$ is a mixing parameter. To select the appropriate Lorentz stucture, 
 $\Pi(q^2)$ is expanded in terms of the electromagnetic charge as:
\begin{equation}
\Pi(q^2)=\Pi^{(0)}(q^2)+e \Pi^{(1)}(q^2,F^{\mu\nu}) + O(e^2).
\end{equation}
The first term $\Pi^{(0)}(q^2)$ is the nucleon propagator which
includes only the CP-even parameters, while the
second term $\Pi^{(1)}(q^2,F^{\mu\nu})$ is the polarization tensor
which may be expanded through Wilson OPE as: $\sum
C_n<0|\bar{q}\Gamma q|0>_{\theta,F}$, where $\Gamma$ is an
arbitrary Lorentz structure and $C_n$ are the Wilson coefficient
functions calculable in  perturbation theory  \cite{SVZ1,IS}. From
this expansion, we keep  only the CP-odd contribution part.
The electromagnetic dependence of these matrix elements
is determined in terms of the magnetic  susceptibilities  $\kappa$,
$\chi $ and $\xi$,  defined as \cite{IS}:
\begin{eqnarray}
<0|\bar{q}\sigma^{\mu\nu} q|0>_F&=& \chi e_q F^{\mu\nu} <0|\bar{q}q|0>
\\ 
g<0|\bar{q}G^{\mu\nu} q|0>_{F}&=& \kappa e_q F^{\mu\nu} <0|\bar{q}q|0> \\ 
2g<0|\bar{q}\tilde{G}^{\mu\nu} q|0>_{F}&=& \xi e_q F^{\mu\nu} <0|\bar{q}q|0>
\end{eqnarray}

 Besides, by considering the anomalous axial current, one obtains the following
$\theta$ dependence of $<0|\bar{q}\Gamma q|0>_{\theta}$ matrix
elements \cite{PR}:

\begin{equation}
m_q <0|\bar{q}\Gamma q|0>_{\theta}= i m_*\theta <0|\bar{q}\Gamma q|0> +  O(m_q^2 )
\end{equation}

where $m_q$ and $m_*$ are respectively the quark and reduced
masses. $O(m_q^2 )$ connection is negligible since $m_\eta >> m_\pi $. 

Putting altogether the above ingredients and after a
straightforward calculation \cite{PR1}, the following expression
of $\Pi^{(1)}(q^2,F^{\mu\nu})$ for the neutron is derived:

\begin{eqnarray}
\Pi(-q^2)&=&-{\bar{\theta}m_* \over
{64\pi^2}}<0|\bar{q}q|0>\{\tilde{F}\sigma,\hat
q\}[\chi(\beta+1)^2(4e_d-e_u) \ln({\Lambda^2\over
-q^2})\nonumber\\ && -4(\beta-1)^2e_d(1+{1\over4}
(2\kappa+\xi))(\ln({-q^2\over \mu_{IR}^2})-1){1\over
-q^2}\nonumber\\ &&-{\xi\over
2}((4\beta^2-4\beta+2)e_d+(3\beta^2+2\beta+1)e_u){1\over
-q^2}...],
\end{eqnarray}

where  $\bar{\theta}=\theta+\theta_q$ is the physical phase and
$\hat q=q_\mu\gamma^\mu$. \\ The QCD expression (2.7) is
confronted to the phenomenological parametrisation
$\Pi^{Phen}$$(-q^2)$ written in terms of the Neutron hadronic
properties. The latter is given by:

\begin{equation}
\Pi^{Phen}(-q^2)=\{\tilde{F}\sigma,\hat q\}
({\lambda^2d_nm_n\over(q^2-m_n^2)^2} +{A\over (q^2-m_n^2)}+...),
\end{equation}

where $m_n$ is the neutron mass, $e_q$ is the quark charge. A and
$\lambda^2$, which  originate from the phenomenological side of
the sum rule, represent respectively a constant of dimension 2 and
the neutron coupling constant to the interpolating current
$\eta(x)$. This coupling is defined via a spinor $v$ as
$<0|\eta(x)|n>=\lambda v e^{\alpha \gamma_5}$.

\section*{Thermal NEDM  sum rules }

The introduction of finite temperature effects may provide
more precision to the phenomenological values of hadronic
observables. Within the framework of QCD sum rules, the
Temperature evolution of the correlation functions manifests itself in the thermal
average of the Wilson operator
expansion\cite{BS,M}. Hence, at relatively low temperature, the system can be regarded as
a non interacting gas of bosons. In this approximatyion, the Thermal dependance of the
vacuum condensates can be written as :
\begin{equation}
<O^i>_T=<O^i>+\int{d^3p\over
2\epsilon(2\pi)^3}<\pi(p)|O^i|\pi(p)>n_B({\epsilon\over T})
\end{equation}
where $\epsilon=\sqrt{p^2+m^2_\pi}$, $n_B={1\over{e^x-1}} $is the
Bose-Einstein distribution and $<O^i>$ is the standard vacuum
condensate (i.e. at T=0). In this approximation, we only kept the pion contributions, since 
in the low temperature region, the effects of heavier resonances 
$(\Gamma= K, \eta,.. etc)$ are dumped by their distibution functions $\sim e^{-
m_\Gamma \over T}$\cite{K}. 
To compute the pion matrix elements, we apply the soft pion theorem given by:
\begin{equation}
<\pi(p)|O^i|\pi(p)>=-{1\over f^2_\pi}<0|[Q^a_5,[Q^a_5,O^i]]|0>+
O({m^2_\pi \over \Lambda^2}),
\end{equation}

where $ \Lambda$ is a hadron scale and $Q^a_5$ is the isovector
axial charge defined by:
\begin{equation}
Q^a_5=\int d^3x \bar{q}(x)\gamma_0\gamma_5{\tau^a\over2}q(x).
\end{equation}
Direct application of the above formula to the quark  and gluon
condensates shows the following features\cite{GL,K}:\\ 
(i) Only $<\bar{q}q>$ is
sensitive to temperature. Its behaviour at finite T is given by:
\begin{equation}
<\bar{q}q>_T\simeq (1-{\varphi(T)\over8})<\bar{q}q>,
\end{equation}
where $\varphi(T)={T^2\over f^2_\pi}B({m_\pi\over T})$ with $B(z)=
{6\over\pi^2}\int_z^\infty dy {\sqrt{y^2-z^2}\over{e^y-1}}$ and
$f_\pi$ is the pion decay constant ($f_\pi\simeq 93 MeV$). The
variation with temperature of the quark condensate $<\bar{q}q>_T$
results in two different asymptotic behaviours, namely:

\begin{equation}
<\bar{q}q>_T \simeq (1- {T^2 \over {8f^2_\pi}}) <\bar{q}q> \quad for \quad {m_\pi \over T}\ll 1 
\end{equation}

\begin{equation}
  <\bar{q}q>_T \simeq (1-{\sqrt {\pi m_{\pi} \over {2T}}}{T^2 \over
{8f^2_\pi}}e^{-m_\pi \over T})<\bar{q}q>  \quad for \quad {m_\pi \over T}\gg 1
\end{equation}

(ii) The gluon condensate is nearly constant at low
temperature and a T dependence occurs only at order $T^8$.

The determination of the ratio ${d_n \over
\bar{\theta}}$ sum rules at non zero temperature is now easily
performed by  applying Borel
operator to both parametrisation of the Neutron correlation function
shown in Eqs. (2.7) and (2.8). Then finite temperature effects
are introduced via the procedure discussed above. Finally, by invoking the
quark-hadron duality, we deduce the final sum rules of the $\bar { \theta}$ induced NEDM at finite temperature:
\begin{equation}
{d_n\over \bar{\theta}}(T)=-{M^2m_* \over 16\pi^2}{1\over
\lambda_n^2(T)M_n(T)}(1-{\varphi(T)\over
8})<\bar{q}q>[4\chi(4e_u-e_d)-{\xi\over 2M^2}(4e_u+8e_d)]e^{M_n^2
\over M^2},
\end{equation}
where M represents the Borel parameter.
The single pole contribution entering the sum rules via the constant
A has been neglected, as suggested in \cite{PR}.\\ 
The value of $\beta$ has been set to 1 in (3.5).
This choice is more appropriate for us since it
suppresses the infrared divergences. The T- evolution of the 
coupling constant  and the mass of the neutron were determined from the thermal nucleon  sum rules \cite{K}.

Within the dilute pion gas approximation, Eletsky has shown
that the contribution induced the
pion-nucleon scattering has to be considered \cite{E}. It enters the nucleon sum rules 
through the coupling constant $g_{\pi NN}$,
whose values lie within the range 13.5-14.3 \cite{PROC}.
\\ \qquad Numerical analysis is performed with the following input
parameters: the Borel mass has been chosen within the values
$M^2=0.55-0.7GeV^2$ which correspond to the optimal range (Borel
window) in the $ d_n\over \bar{\theta}$ sum rule at $T=0$
\cite{PR1}. For the $\chi$ and $\xi$ susceptibilities we take
$\chi=-5.7\pm 0.6 GeV^{-2}$ \cite{BK} and $\xi=-0.74\pm 0.2$
\cite{KW}. As to the vacuum condensates appearing in (3.5), we use
their standard values \cite{SVZ}.

\section*{Analysis and Conclusion}

We have established the relation
between the NEDM and $ \bar{\theta} $ angle at non zero
temperature from QCD sum rules. We find that the behaviour of the ratio${ d_ n \over\bar{\theta}}$
 is connected to the thermal evolution  of the pion parameters
$f_\pi$, $m_\pi$ and of $g_{\pi NN}$.\\
 By analysing the ratio  as a function of T in the region of validity of thermal sum-rules $[0,T_c]$, we learn that  $\mid{ d_ n \over \bar{\theta}}\mid$  decreases smoothly
with T (about 16$\%$ variation for temperature values up to 200
MeV) but survives at finite temperature. This means that either the NEDM value decreases or
$\bar{\theta}$ increases. Consequently, for a fixed value of
$\bar{\theta}$ the NEDM decreases but it does not exhibit any
critical behaviour. Furthermore, if we start from a non vanishing
$ \bar{\theta} $ value at $T=0$, it is not possible to remove it
at finite temperature. We also note that $ \mid{d_n\over
\bar{\theta}}\mid$ grows as
 $M^2$ or $\chi$ susceptibility increases.
It also grows with quark condensate rising. However this ratio is
insensitive to both the $\xi$ susceptibility and the coupling
constant $g_{\pi NN}$. We notice that for higher temperatures, the
analysis of $\mid{d_n\over \bar{\theta}}\mid=f({T\over T_c})$ exhibits a
brutal increase justified by the fact that for temperatures beyond
 the critical value $T_c$, at which the chiral symmetry is restored,
the constants $f_\pi$ and $g_{\pi NN}$ become zero and
consequently from  Eq(3.5) the ratio $ {d_n\over \bar{\theta}}$
behaves as a non vanishing constant. The large difference between
the values of the ratio for $T<T_c$ and $T>T_c$ may be a consequence
of the fact that other  contributions
to the the spectral function have ben neglected, like the scattering process $ N+ \pi \to \Delta $. These contributions
 which are of the order $T^4$, are negligible in the low temperture region but become 
substantial for $T\ge T_c$.
Moreover, this difference may also originate from the use of soft
pion approximation which is valid essentially for low $T$ ($T<
T_c$). Therefore it is clear from this qualitative 
analysis, which is based on the soft pion approximation, that 
temperature does not play a fundamental role in the suppression of
the undesired $\theta$-term and hence the broken CP symmetry is
not restored \cite{MS}. Indeed, some exact symmetries can be broken by
increasing temperature \cite{W,MS}. The symmetry non restoration
phenomenon, which means that a broken symmetry at T=0 remains
broken even at high temperature, is essential for discrete
symmetries, CP symmetry in particular. Indeed, the symmetry non
restoration allows us to avoid wall domains inherited after the
phase transition \cite{ZKO} and to explain the baryogenesis
phenomenon in cosmology \cite{S}. Furthermore, it can be very
useful for solving the monopole problem in grand unified theories
\cite{DMS}.\\ \\

\centerline {\bf AKNOWLEDGMENTS}

The author is  deeply grateful to Prof. Brigitte Hiller and to the organization Committe for the invitation
to the Second International Conference on Hadron Physics. 
He also wishes to thank Prof. Joao Providencia
for his hospitality during the visit to CPT at Coimbra.
\\ This work is supported by  the convention de cooperation between
 CNRST/ICCTI 681.02/CNR.

\end{document}